\documentclass[letterpaper]{jpsj3}
\usepackage{color}

\title{Computing Chemical Potential using the Phase Space Multi-histogram Method}

\author{\name{Hitomi} \surname{Nomura}\thanks{E-mail address: tmc27563@st.yamagata-u.ac.jp}, \name{Tomonori \surname{Koda}}, \name{Akihiro \surname{Nishioka}}, \name{Ken \surname{Miyata}} 
}
\inst{\address{Graduate School of Science and Engineering, Yamagata University, 4-3-16 Jonan, Yonezawa, Yamagata 992-8510, Japan} 
}

\abst{We present a new simulation method to calculate the free energy and the chemical potential of hard particle systems. The method relies on the introduction of a parameter dependent potential to smoothly transform between the hard particle system and the corresponding ideal gas. We applied the method to study the phase transition behavior of monodispersed infinitely thin square platelets. First, we equilibrated the square platelet system for different reduced pressures with a usual isobaric Monte Carlo (MC) simulation and obtained a reduced pressure-chemical potential plot. Then we introduce the parametrized potential to interpolate the system between the ideal gas and the hard particles. After selecting the potential, we performed isochoric MC runs, ranging from the ideal gas to the hard particle limit. Through an iterative procedure, we compute the free energy and the chemical potential of the square platelet system by evaluating the volume of the phase space attributed to the hard particles, and then we find the coexistence pressure of the system.  Our method provides an intuitive approach to investigate the phase transitions of hard particle systems.}

\kword{multi-canonical method, isotropic-nematic phase transition, isobaric and isochoric Monte Carlo simulations, phase space multi-histogram method, chemical potential}

\begin{document}
\maketitle

\section{Introduction}
Since Onsager's pioneering work~\cite{o49}, it has been studied the phase transition of rod-shaped and disc-shaped hard particle systems using simulations with excluded volume effects. For example, the isotropic-nematic-smectic \textit{A} phase transition was investigated by Frenkel et al.~\cite{f88} using an isobaric MC of hard spherocylinders. Frenkel and Mulder~\cite{f85} and Veerman and Frenkel~\cite{v92} observed various phase transitions of disk-shaped hard particles modeled as oblate ellipsoids and cut spheres respectively. The isotropic-nematic phase transition of monodispersed infinitely thin disks was first studied by Eppenga and Frenkel~\cite{e84}, and then the nematic-columbar phase transition of those was observed by  Bates and Frenkel~\cite{f98}. 

In this paper, we introduce a new simulation method which we shall call the "phase space multi-histogram" (PSMH) method to compute chemical potential of hard particles.  The method is a variant of the multi-canonical ensemble method~\cite{bn91, bn92}. As an illustration of the PSMH method, we compute the variations in the chemical potentials for nematic and isotropic phases and detect the isotropic-nematic phase transition point for the square platelet system examined by Bates in 1999~\cite{mb99}. 

Bates' computer simulation work~\cite{mb99} was motivated from an experimental result by van der Kooij and Lekkerkerker who synthesized suspensions of sterically stabilized gibbsite ($Al(OH)_{3}$) platelets and observed the isotropic-nematic phase transition in a critical range of concentration of the experimental system\cite{kl98}. Examining the nature of their experimental system, Bates studied the isotropic-nematic phase transition of monodisperse infinitely thin platelet systems using grand canonical ($\mu VT$ ensemble) Monte Carlo (MC) simulations. It was found that the shape of a platelet is essential to determine its isotropic-nematic phase transition behavior. In particular,  the isotropic and nematic coexistence reduced densities of the isotropic and nematic  phases for the system of square platelets were in the range of 3.1-3.4 due to hysterisis.

The organization of this paper is the followings. In Section \ref{psmh}, the PSMH method is explained after briefly reviewing the canonical and multi-canonical ensemble MC simulations.
The method can be used to calculate the free energy of hard repulsive particles by interpolating from the ideal gas state. In Section \ref{eg}, the PSMH method is used for the detection of the isotropic (I) - nematic (N) phase transition of monodispersed infinitely thin square platelets. Finally, the conclusion of the present study is given in Section \ref{con}.  
\begin{figure}[h]
\begin{center}$
\begin{array}{cc}
\includegraphics[scale=0.3]{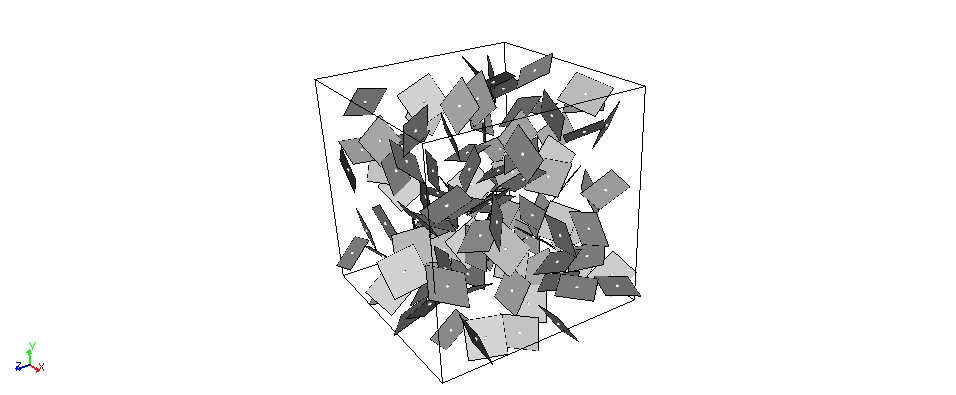} &\\
\includegraphics[scale=0.32]{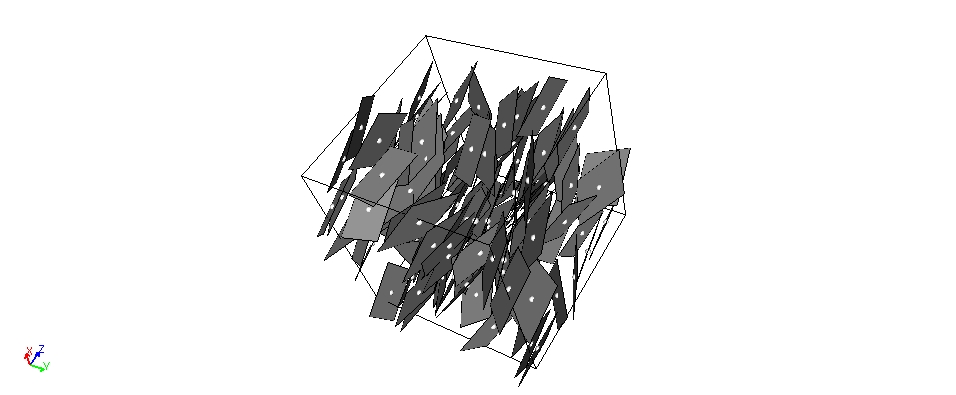}
\end{array}$
\end{center}
\caption{(top) Isotropic phase and (bottom) Nematic phase of the system of 120 square platelets. The white balls indicate the centers of mass of the square platelets.}
\label{fig:fig1}
\end{figure}

\section{The PSMH Method}\label{psmh}
\subsection{Passing from Canonical to Multi-canonical}
Consider a non-isolated macroscopic system in contact with a heat reservior. Under the constraint that the number of particles $N$, volume $V$ and temperature $T$ are constant, the probability $P_{NVT}$ that the system is in microstates with energy $E$ is proportional to $w(E)\exp (-\beta E)$. Here, $w(E)$ is the number of microstates in the same energy $E$, and $\beta$ is  the inverse temperature defined by $\beta = 1/(k_{B}T)$, where $k_{B}$ is Bolzmann's constant. The ensemble defined by $P_{NVT}$ is called canonical. It is known that the canonical method has the limitation of not being able to escape from a metastable state, and the multi-canonical method was introduced by Breg and Neuhaus~\cite{bn91, bn92} to remedy the situation. The multi-canonical probability $P_{m}$ has a uniform probability distribution such that $P_{m}$ is proportional to $\varrho(E)\exp (-\beta E_{m})$, where $\varrho(E)$ is the density of states, and $E_{m}$ is the multi-canonical energy~\cite{m00}. The multi-canonical ensemble MC simulation is carried out on the basis of the Metropolis algorithm~\cite{m53} with the re-weighting histogram technique to study phase transitions~\cite{fs88}.

\subsection{Introducing a Parameterized Potential}
A key ingredient of the PSMH method is a parameter dependent potential $U_{h}$ to smoothly transform between the hard particle system and the corresponding ideal gas state. Recall that the classical partition function of the potential energy of a system, say $U$, is given by
\begin{equation}\label{eqn:zb1}
Z(\beta)=\int dU\varrho(U)\exp(-\beta U),
\end{equation}
and once one knows all information about the density of states $\varrho(U)$, the partition function $Z(\beta)$ is computable for any $\beta$ multi-canonically by the re-weighting histogram method~\cite{bn91, bn92, fs88}. Consider a hard particle system with a pairwise repulsive potential energy such that $\phi_{ij} = \infty$ if the $i$- and $j$-th particles overlap, and $\phi_{ij} = 0$ if they do not. Then the total potential energy of the system is given by $U_{h} \equiv \sum_{i<j}\phi_{ij}$. By Eqn. (\ref{eqn:zb1}), the partition function of the potential energy $U_{h}$ is written as
\begin{equation}\label{eqn:zb2}
Z(\beta)=\int dU_{h}\varrho(U_{h})\exp(-\beta U_{h})=\int dU_{h}w(U_{h})\delta (U_{h}-U_{0})\exp(-\beta U_{h})=w(U_{0}), 
\end{equation}
where $\delta$ is the Dirac delta function, and $U_{0}=0$.
Since the partition function $Z(\beta)$ does not depend on $\beta$ any more, we may write $Z(\beta) \equiv Z$.

In order to compute $w(0)$, we shall introduce a parameter dependent potential, $U_{h}'$ defined by
\begin{equation}\label{eqn:U}
U_{h}' \equiv \frac{hn_{c}}{\beta},
\end{equation}
where $h$ is an interpolation constant, and $n_{c}$ is the number of pair intersections such that
\begin{equation}\label{eqn:nc}
n_{c}=\sum_{i < j}
\left\{
\begin{array}{ll}
1 & \mbox{if the i-th and j-th particles overlap} ,\\
0 & \mbox{otherwise}. 
\end{array}
\right.\\
\end{equation}
In particular, for a distribution $\exp(-\beta U_{h}')$, $n_{c}$ is fluctuating when $h$ is constant, and we have the following relationship
\begin{equation}\label{eqn:h}
\left\{
\begin{array}{ll}
 \mbox{when }h \rightarrow \infty , & n_{c} \rightarrow 0 \mbox{ (hard body model)} ,\\
  \mbox{when }h \rightarrow 0, & n_{c} \ne 0  \mbox{ (ideal gas)} .
\end{array}
\right.
\end{equation}
In reality, the parameter dependent potential $U_{h}'$ behaves as the potential $U_{h}$ as $h$ goes to infinity
\begin{equation}\label{eqn:Uhlim}
\lim_{h \rightarrow \infty}U_{h}' =U_{h},
\end{equation}
and the partition function $Z'(\beta)$ of the parameter dependent potential $U_{h}'$ is
\begin{equation}\label{eqn:zb3}
Z'(\beta)=\int dn_{c}w(n_{c})\exp(-hn_{c}).
\end{equation}
Writing $\beta U_{h}$ as the product of an interpolation constant $h$ and the number of pair intersections $n_{c}$ enables one to describe the model of an ideal gas which allows non-zero pair intersections. 

The PSMH method uses the re-weighting multi-histogram technique~\cite{fs88}. The PSMH method is also similar to the multi-canonical method~\cite{bn91, bn92} since it changes $h$ which is equivalent to changing $\beta$, but it is different in the following respect. The PSMH method adopts the multi-canonical method to obtain $w(0)$; however, we do not actually perform the multi-canonical method for the potential $U_{h \rightarrow \infty}$ whose partition function $Z$ is independent on $\beta$ as seen in Eqn. (\ref{eqn:zb2}). The PSMH method is a variant of the multi-canonical method~\cite{bn91, bn92} to study phase transitions of hard particle systems with the re-weighting technique~\cite{fs88}.

\subsection{The PSMH method: Procedures and Calculation}
First, we perform a usual isobaric MC simulation for hard particles. It corresponds to the limit of $h \rightarrow \infty$ in  Eqns. (\ref{eqn:U}), (\ref{eqn:nc}). After equilibrating, we switch the simulation scheme to an isochoric one and remove the constraint of non-overlapping of hard particles, in other words, we perform an isochoric MC simulation with the Boltzmann weight factor $\exp(-\beta U_{h})$. We call the latter scheme as NVh ensemble MC simulations. In the NVh MC runs, the number of particle pair intersections $n_{c}$ fluctuates. Let $w(n_{c})$ be the number of microstates such that there are $n_{c}$ pair contacts in each microstate. From the NVh MC simulation data, we construct the apparent distribution $\psi_{h}(n_{c})$ of $n_{c}$ for a given $h$. $\psi_{h}(n_{c})$ obeys the normalization condition $\sum_{n_{c}}\psi_{h}(n_{c})=1$ ("Histogram"). Let $C_{h}$ be a normalization constant and $h$ be incremented by $\Delta h$. Then
\begin{equation}\label{eqn:hb}
\psi_{h}(n_{c})=C_{h}w(n_{c})\exp(-\beta U_{h}'),
\end{equation}
\begin{equation}\label{eqn:hp}
\psi_{h+\Delta h}(n_{c})=\frac{C_{h+\Delta h}}{C_{h}}\psi_{h}(n_{c})\exp(-\Delta hn_{c}).
\end{equation}

Since $n_{c}$ is an integer, we call it using the letters $j, k, l$ simply. A straightforward calculation gives us 
\begin{equation}\label{eqn:wb}
\ln w(k)=\ln \psi_{h=0}(k) - \ln C_{0}, k=j, j-1, j-2, \cdots , l, l < j
\end{equation}
when $h=0$, and 
\begin{equation}\label{eqn:wi}
\ln w(l-1)= \ln \psi_{h}(l-1) - \ln \psi_{h}(l)+ \ln w(l) + h
\end{equation}
when $h \ne 0$. Here, $C_{0}^{-1} = \sum_{n_{c}}w(n_{c})$.

Assume that the global minimum value of  $\psi_{h}(n_{c})$ is attained, and then computation of the free energy $F$ and chemical potential $\mu$ of the $N$ hard particles are carried out via 

\begin{equation}\label{eqn:F}
\beta F \approx - \ln w(0) + \ln C_{0}
\end{equation}
and 

\begin{equation}\label{eqn:M}
\beta \mu = \frac{\beta(F+pV)}{N} = \frac{\beta F}{N}  + \frac{p^{\ast}}{N}(\frac{L}{D})^3,
\end{equation}
where $p, V, L, D, p^{\ast}$ are a pressure, the volume of a periodic simulation box, the side length of the simulation box, the unit length in simulation, and a reduced pressure
\begin{equation}\label{eqn:ul}
p^{\ast}=D^{3}p\beta.
\end{equation}

The estimation of $C_{0}^{-1}$ is given in Appendix~\ref{app}, and we set the thermal de Broglie wavelength as $\Lambda = 1$ in simulation. 

\section{Illustration of the PSMH method}\label{eg}
\subsection{Model system}
Our model system is a monodispersed square platelet system. The system is composed of $N=120$ square platelets of equal area $A$. The unit length $D$ in Eqn. (\ref{eqn:ul}) is set by $D=\frac{\sqrt{A}}{2}$. We performed isobaric ($NP$ ensemble) MC simulations. The initial configuration of the system was a dilute aligned crystal with a cubic periodic condition, and it was relaxed at $p^{\ast}=0.1$ to become  an isotropic phase. We isotropically compressed it (i.e. the shape of the periodic simulation box was always cubic) at the reduced pressure ranged in $p^{\ast}=0.1$-$2.0$, where $\beta$ is the inverse temperature. After that, we expand it at the reduced pressure in the same range. For compression (or expansion), an initial structure was an equilibrium structure obtained by a previous simulation. At each pressure, the system was equilibrated to compute statistical quantities. The isotropic-nematic phase transition of the system is driven by an excluded volume repulsion (i.e. colloidal interactions which prohibit any overlap of platelets.)

Following Bates' notation,~\cite{mb99} we analogously define a number density $\tilde{\rho}=\frac{N\tilde{\sigma}^3}{V}$ with a scaled diameter $\tilde{\sigma}=\sqrt{\frac{4A}{\pi}}$. Let $\widetilde{\rho_{NI}}$ and $\widetilde{\rho_{IN}}$ be the number coexistence densities along two branches, that is, one from the nematic phase to the isotropic phase ($NI$ branch) and the other from the isotropic phase to the nematic phase ($IN$   
branch) of the system of square platelets due to hysteresis. In simulation, the IN branch is obtained by compression, and the NI branch is done by expansion.

A nematic order parameter matrix $\mathcal{S}$ is a matirx quantity which measures the isotropic-nematic phase transition and is given by 
\begin{equation}\label{eqn:S}
\mathcal{S} \equiv (\mathcal{S}_{ij})=\frac{1}{N}\sum_{k=1}^{N}\{\frac{3}{2}(r_{k}\cdot e_{i})(r_{k} \cdot e_{j})-\frac{1}{2}\delta_{ij} \}
\end{equation}
when $N$ is the fixed number of the square platelets in the system. $e_{i}, e_{j}$ are unit vectors along the lab coordinates and $r_{k}$ is a unit vector along the $k$th molecule's symmetric axis here. Since each entry $(\mathcal{S}_{ij})$ is  a real number for any $i, j$, and $(r_{k}\cdot e_{i})(r_{k} \cdot e_{j}), \delta_{ij}$ are symmetric with respect of $i$ and $j$, $\mathcal{S}$ is a real symmetric matrix. Therefore, $\mathcal{S}$ is diagonalizable. The largest eigenvalue of the diagonalized $\mathcal{S}$ is called an order parameter denoted by $\mathrm{S}$ and its corresponding eigenvector is called a director. A model system with a nematic behavior shows a nematic phase with an orientational order (i.e. a director without any positional order.)

\subsection{Results and Discussion}
\begin{figure}
\begin{center}
\includegraphics[scale=0.65]{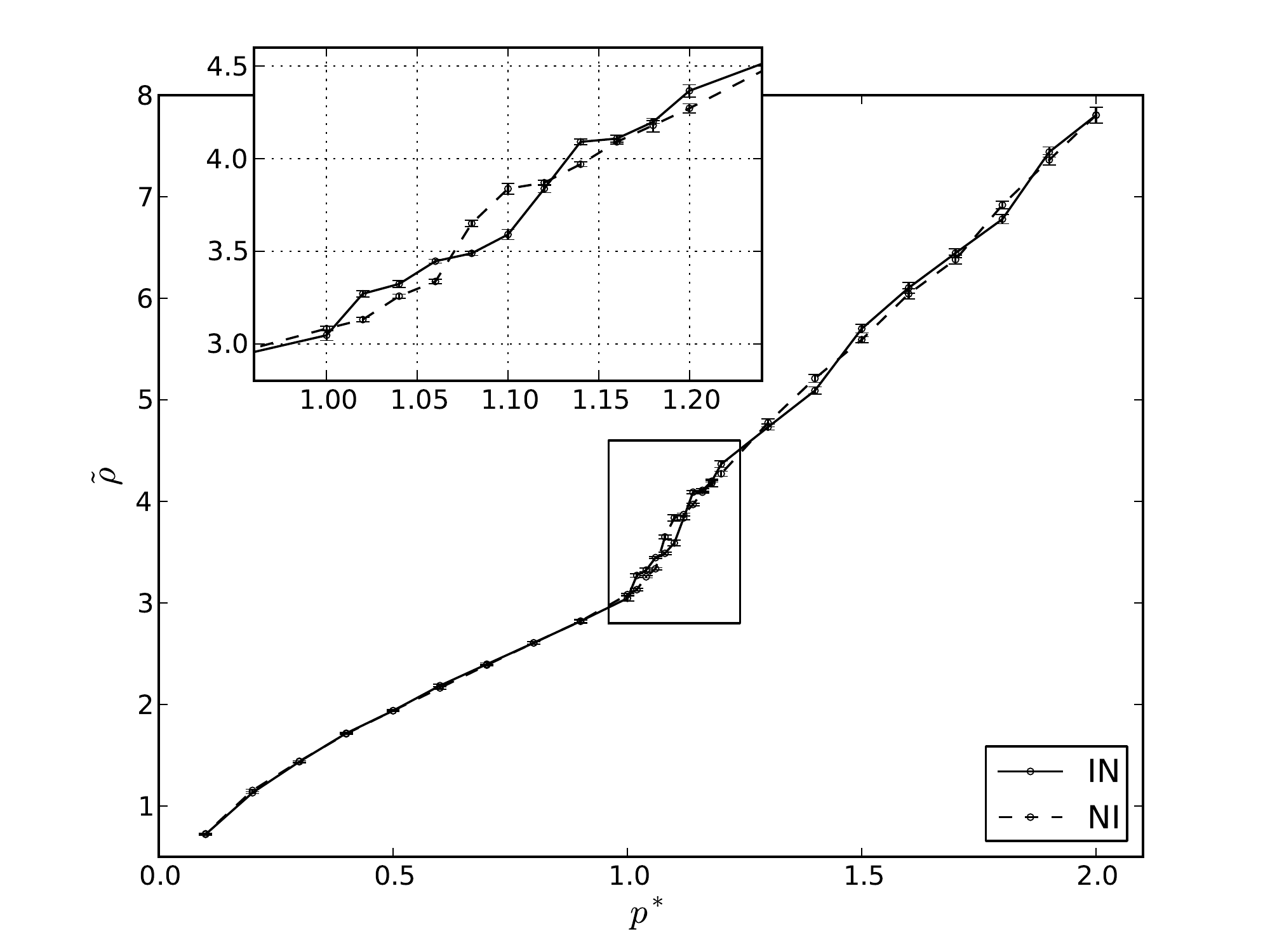}\\
\end{center}
\caption{$p^{\ast}$ vs. Number density $\tilde{\rho}$. $IN$ branch means a phase transition from isotropic to nematic, and vice versa. A weak hysteresis was observed around $p^{\ast}=1.1$. A vertical error bar on each data point indicates the standard error of the mean. For a guide to the eyes, the lines between the data are drawn. }
\label{fig:fig2}
\end{figure}

\begin{figure}
\begin{center}
\includegraphics[scale=0.65]{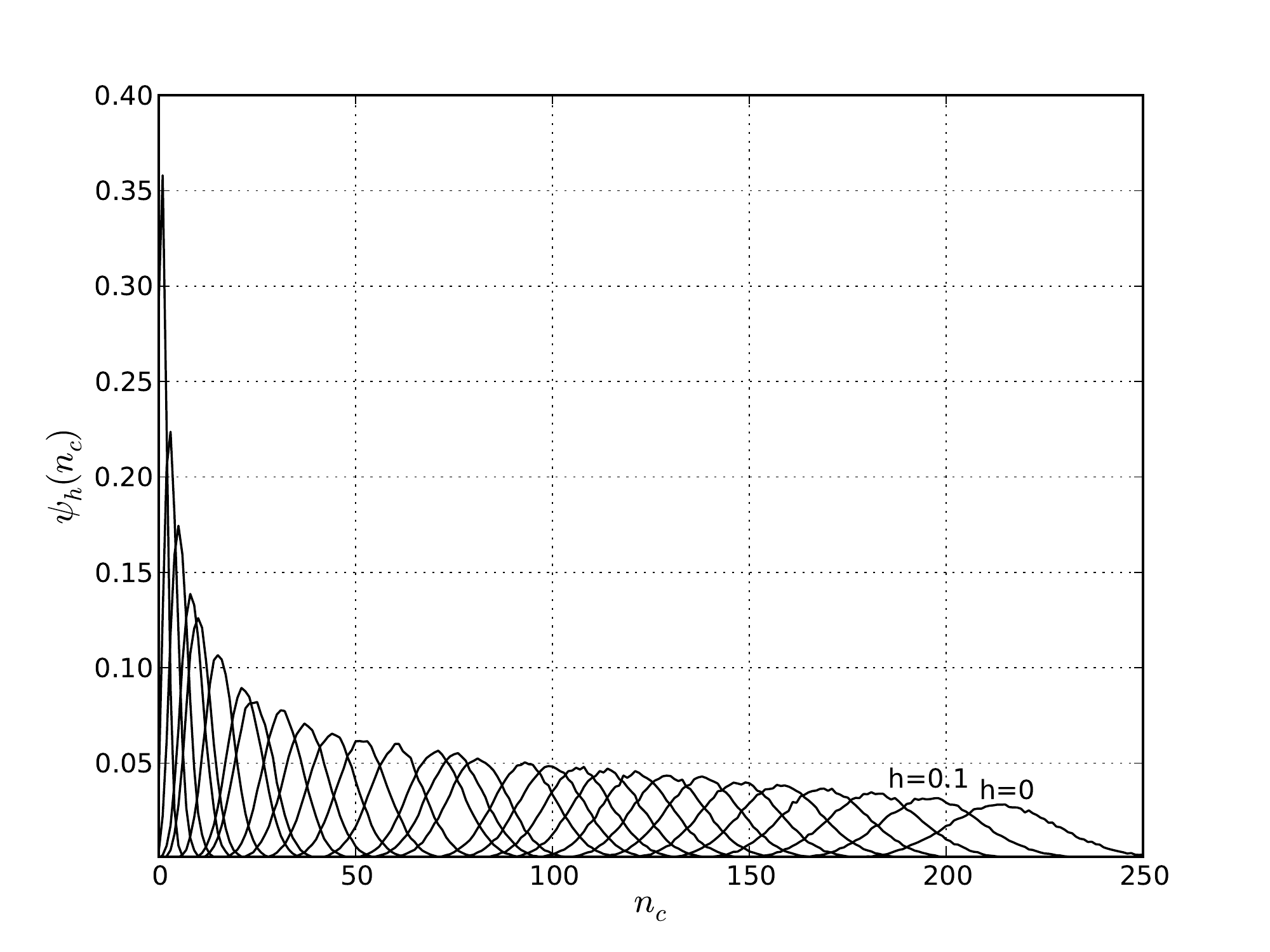}
\end{center}
\caption{A plot of histograms $\psi_{h}(n_{c})$ for $h$ ranged in $[0, 6]$. The $NI$ branch at $p^{\ast}=0.8$ was taken. When $h=0$, the maximum of $\psi_{h}(n_{c})$ was $0.02842$ at $n_{c}=215$.}
\label{fig:fig5}
\end{figure}

\begin{figure}
\begin{center}
\includegraphics[scale=0.65]{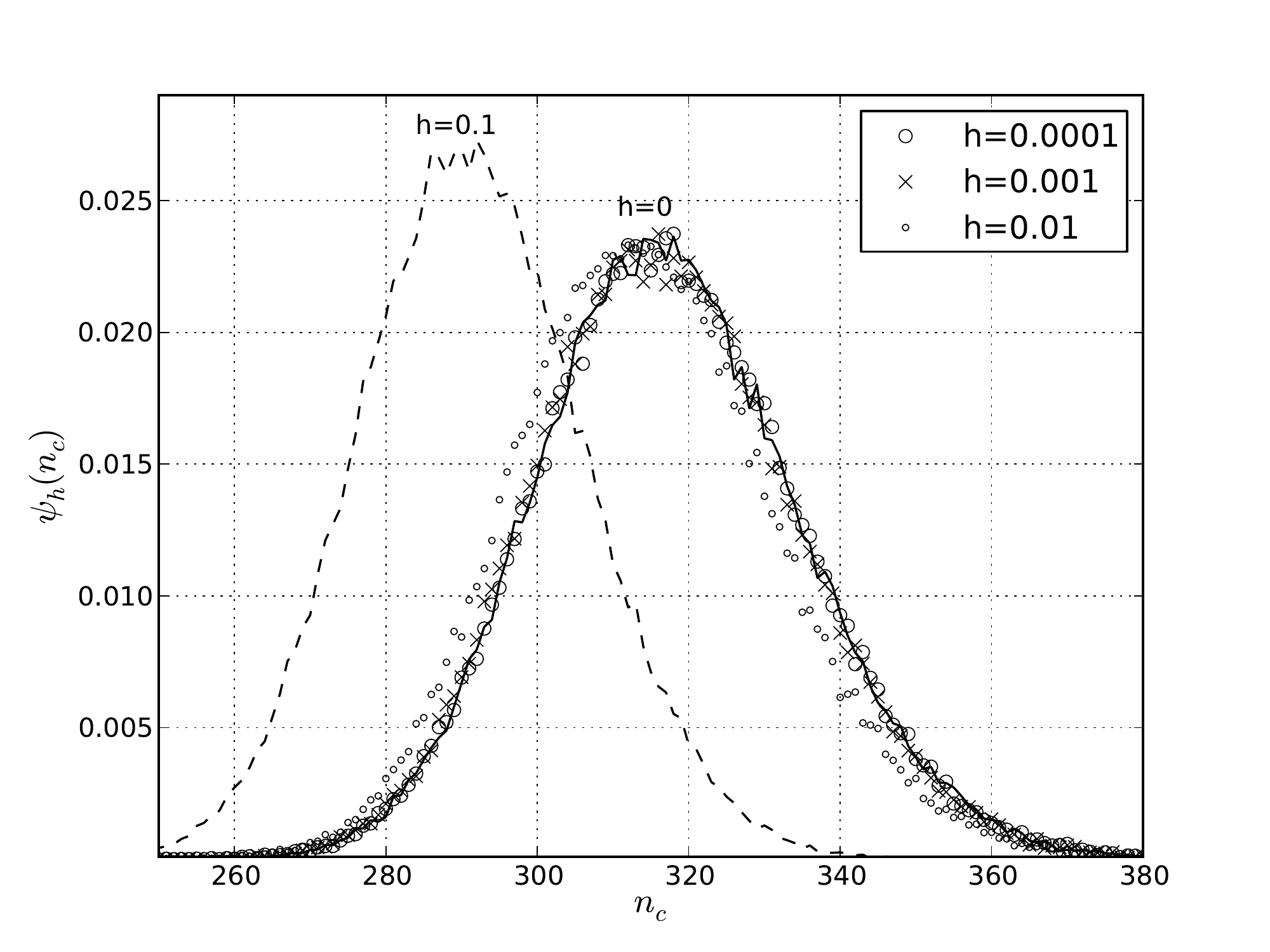}
\end{center}
\caption{A plot of histograms $\psi_{h}(n_{c})$ for $h=0, 0.01, 0.001, 0.0001, 0.1$.}
\label{fig:fig111}
\end{figure}

\begin{figure}[h]
\begin{center}$
\begin{array}{cc}
\includegraphics[scale=0.5]{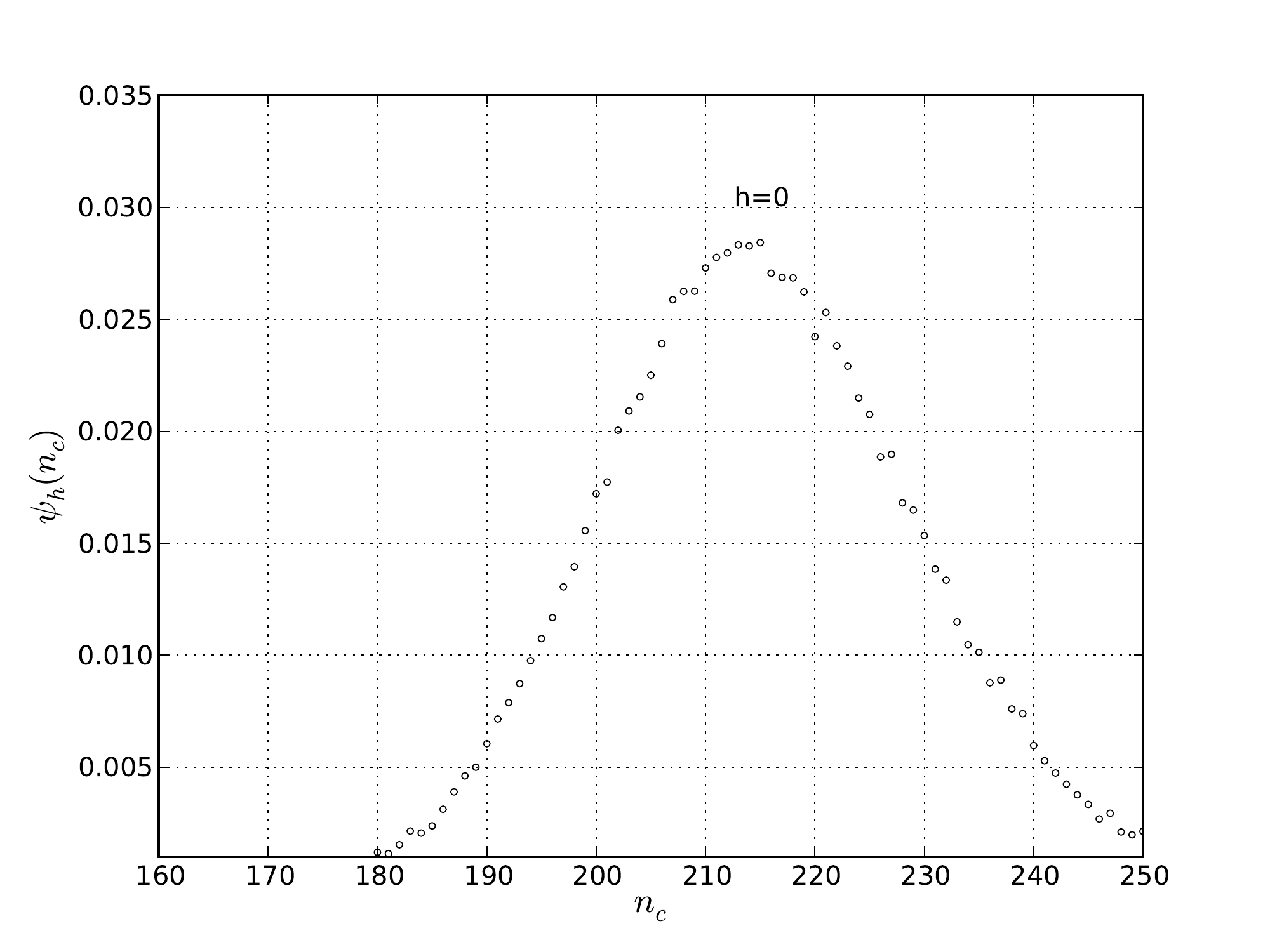} &\\
\includegraphics[scale=0.5]{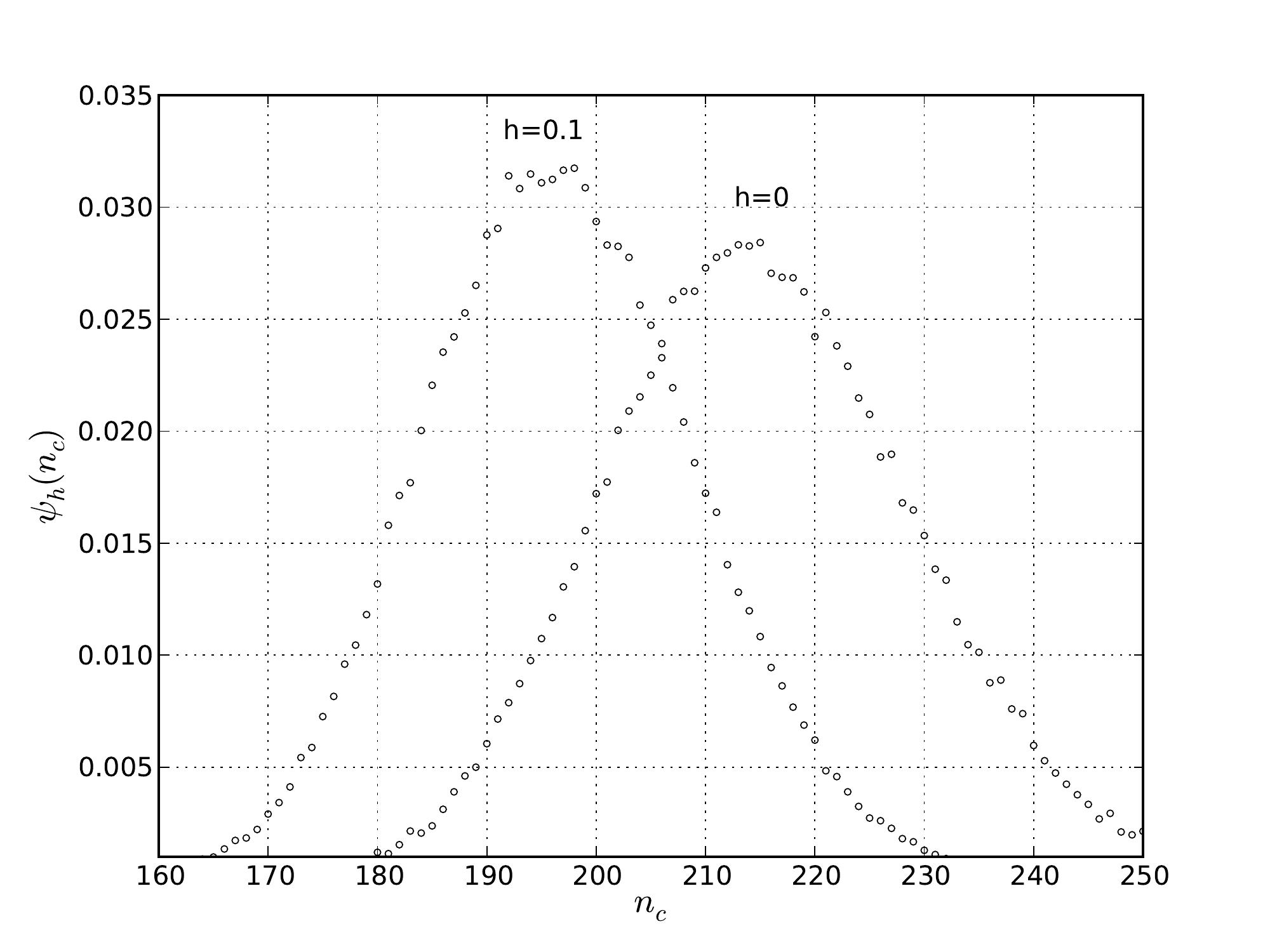}
\end{array}$
\end{center}
\caption{(top) A histogram $\psi_{h}(n_{c})$ for $h=0$ and (bottom) Two histograms $\psi_{h}(n_{c})$ when $h=0,~0.1$. The $NI$ branch at $p^{\ast}=0.8$ was used. $j=319$ and $l=201$ were taken when $h=0$. The nieghboring histograms overlapped within full width at half maximum.}
\label{fig11}
\end{figure}

\begin{figure}
\begin{center}
\includegraphics[scale=0.65]{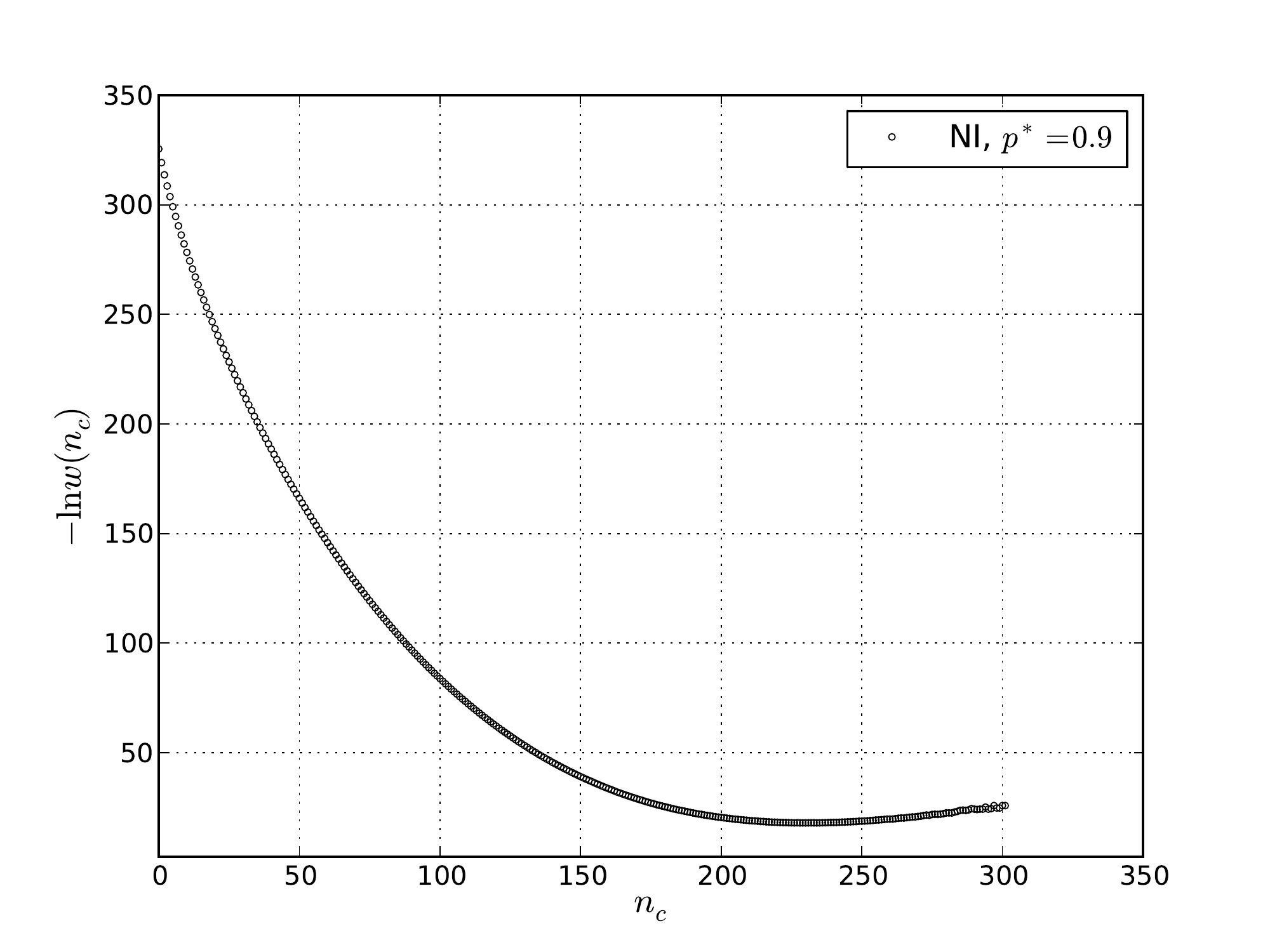}
\end{center}
\caption{$n_{c}$ vs. $-\ln w(n_{c})$ for the NI branch at $p^{\ast}=0.9$.}
\label{fig4}
\end{figure}

\begin{figure}
\begin{center}
\includegraphics[scale=0.65]{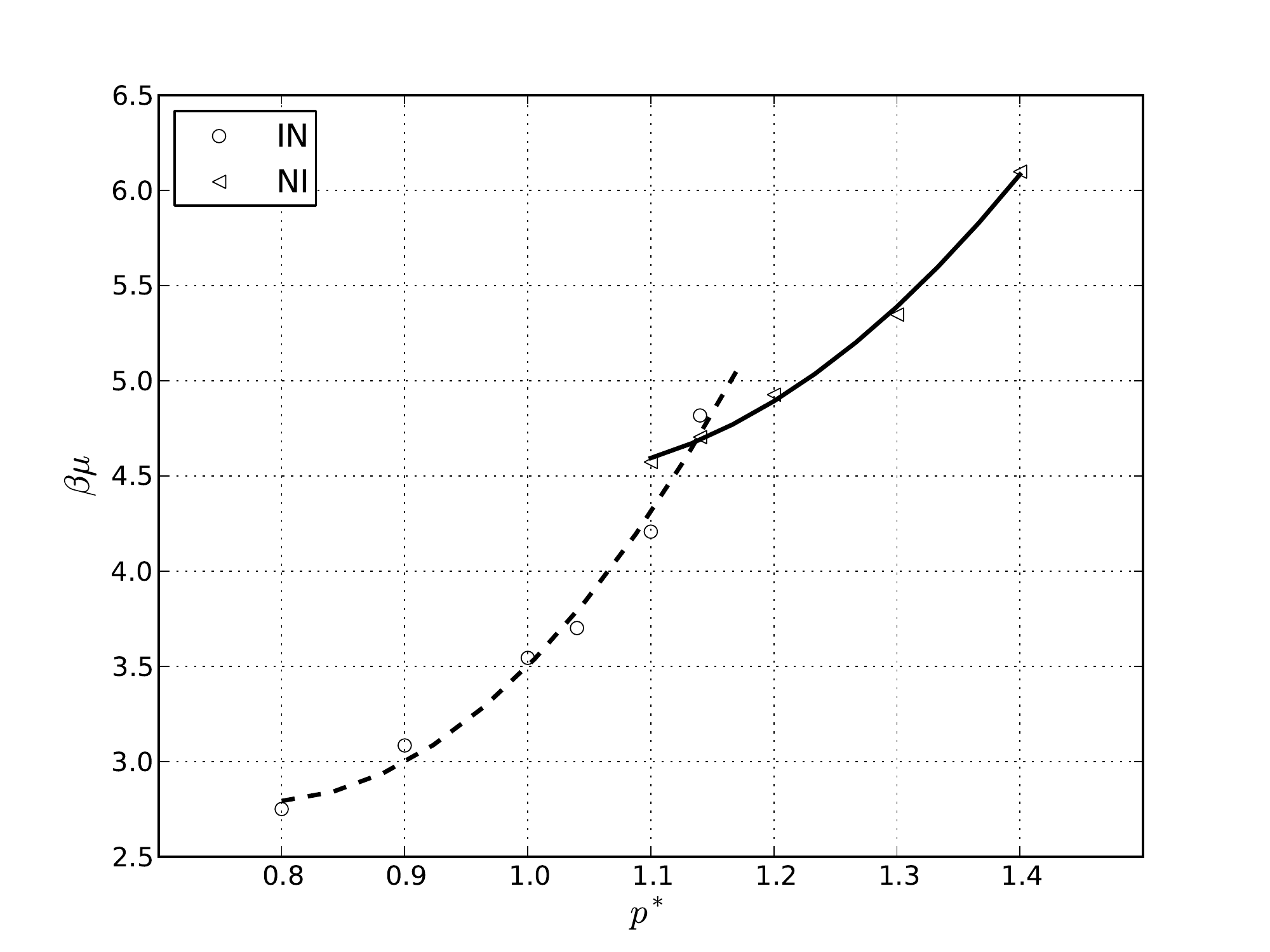}
\end{center}
\caption{$p^{\ast}$ vs. chemical potential curves for $NI$ and $IN$ branches. The solid curve ($NI$ branch) represents a polynomial fit to the $NI$ data, and the dotted curve ($IN$ branch) is a polynomial fit to the $IN$ data. The phase transition was observed at  $p^{\ast} = 1.13~(\beta\mu =4.61)$.}
\label{fig:fig6}
\end{figure}

In FIG.~\ref{fig:fig1}, we displayed typical isotropic ($\mathrm{S} \approx 0.1$)  and nematic ($\mathrm{S} \approx 0.7$) phase configurations of a square platelet system.  Due to the finite size of the system, hysteresis in density was observed around $p^{\ast}=1.1$ as pressure was changed. (See FIG.~\ref{fig:fig2}.) In FIG. \ref{fig:fig5}, a plot of histograms $\psi_{h}(n_{c})$ for each $h$ ranged in $[0, 6]$ was shown. The $NI$ branch at $p^{\ast}=0.8$ was selected initially. FIG.~\ref{fig:fig111} shows the behavior of histograms when each h is on the interval $[0, 0.1]$. The histograms for $h=0.1$ and $0$ are shown in dotted and solid curves. We found the histogram for $h \rightarrow 0$ converging to that for $h=0$. (See FIG.~\ref{fig:fig111}.) The PSMH method worked in the square platelet system since multi-histograms could generate overlapped histograms within full width at half maximum. (See FIG. \ref{fig11}.) The $n_{c}$ vs. $-\ln w(n_{c})$ plot for the NI branch at $p^{\ast}=0.9$ were shown in FIG. \ref{fig4}. When $n_{c}$ was big, $-\ln w(n_{c})$ took a small value. $-\ln w(n_{c})$ attained the global minimum at $n_{c}=234$, and we found $-\ln w(n_{c})\approx 325$ when $n_{c}=0$. In FIG. \ref{fig:fig6}, $p^{\ast}$ vs. chemical potential curves for $IN$ and $NI$ branches was shown. The phase transition was observed at  $p^{\ast} = 1.13~(\beta\mu =4.61)$. The order parameter was $\mathrm{S} \approx 0.7$ when the number coexistence densities of the system of square platelets were  $\widetilde{\rho_{NI}}\approx 3.85$ and $\widetilde{\rho_{IN}} \approx 3.87$ (at $p^{\ast}=1.13$). This was comparable to Bates' $(\rho_{NI}, \rho_{IN}) \approx (3.1, 3.4)$~\cite{mb99} when taking into account of the influence of systematic errors. In colloidal suspensions of $Al(OH)_{3}$ platelets, the number coexistence densities $\widetilde{\rho_{IN}}, \widetilde{\rho_{NI}}$ and the difference between those densities,  $\widetilde{\rho_{IN}} - \widetilde{\rho_{NI}}$ were known to be very small as an experimental fact.~\cite{mb99}. Although, in our simulation, the number coexistence densities were relatively high, but the difference was only $\widetilde{\rho_{IN}} - \widetilde{\rho_{NI}}=0.02$. 

In Bates' work~\cite{mb99}, he used a different expression:
\begin{equation}\label{eqn:bmpb}
\beta\mu_{Bates} = \beta\mu + (\log \frac{4\pi V}{D^3}).
\end{equation}
Our simulation result in chemical potential at the isotropic-nematic phase transition was $\beta\mu=4.61$, and Bates' was $\beta\mu_{Bates}=7.3$. By using Eqn. (\ref{eqn:bmpb}),  we get 
\begin{equation}\label{eqn:diffd5}
\beta\mu=\beta\mu_{Bates}-\log(4\pi V)\approx 4.1
\end{equation}
when $V$ is the $5\times5\times5$ simulation box, and
\begin{equation}\label{eqn:diffd8}
\beta\mu=\beta\mu_{Bates}-\log(4\pi V)\approx 3.5
\end{equation}
when $V$ is the $8\times8\times8$ simulation box. In either case, our simulation result was comparable to Bates'. We did not studied the effect of the shape of the simulation box, but our result combined with Betes'~\cite{mb99} suggested that there was the size dependence of the free energy and chemical potential of the monodispersed infinitely thin square platelet system.

While having had such a successful example of the PSMH method, the method has its limitation for some hard particle systems where particle insertion is rejected. For example, the PSMH method did not work in the sphere system to study a phase transition from a solid state to a liquid state~\cite{nk11}. This indicates that the PSMH method works well in hard particle systems where particle insertion is possible.

\section{Conclusions}\label{con}
We introduced the phase space multi-histogram (PSMH) method to calculate the free energy and the chemical potential of hard particle systems. The most essential part of the PSMH method is introducing the parameter dependent potential $U_{h}$ to smoothly transform between the hard particle system ($h \rightarrow \infty$) and the corresponding ideal gas limit ($h \rightarrow 0$), and a basic idea of the PSMH method is shrinking the phase space of the ideal gas state to that of the hard particle system using the potentials $U_{h}$ parameterized by $h$.

We investigated the isotropic-nematic phase transition of the system of monodispersed infinitely thin square platelets by applying the PSMH method. After equilibrating the square platelet system for different reduced pressures with isobaric MC runs, we obtained a reduced pressure-chemical potential plot. Next, we introduced the parametrized potential to interpolate the system from the ideal gas state to the hard particle system. Following that, we selected the potential, and then we performed isochoric MC runs with varying from the ideal gas state to the hard particle limit. By the induction procedure of the PSMH, we computed the free energy and the chemical potential of the square platelet system by evaluating the volume of the phase space attributed to the hard particles, and then we find the coexistence pressure of the system. To detect an isotropic-nematic phase transition point for the square platelet system, we calculated the variations in chemical potentials for nematic and isotropic phases.  For the system, the obtained number coexistence densities and the chemical potential at the isotropic-nematic phase transition point closely matched up with Bates' results.~\cite{mb99} The PSMH method works well only if multi-histograms are generated so that the neighboring histograms overlap sufficiently.

This is a pilot study, and we will continue to develop the PSMH method. In particular, we will investigate the behavior of the system with the choice of $h$, the dependance of the size of platelets and the influence of a simulation box geometry.  The next step in this work is to study the isotropic-biaxial nematic phase transition of monodispersed infinitely thin irregular platelet systems which has as of yet not been detected.

\begin{acknowledgment}
I wish to thank Jun-ichi~Takimoto and Sathish~K. Sukumaran for their helpful discussions. Special thanks go to the an anonymous referee for providing me with constructive comments and suggestions.
\end{acknowledgment}

\appendix
\section{Estimation of $C_{0}^{-1}$}\label{app}
In order to compute chemical potentials, we need to estimate $C_{0}^{-1}$ for a monodispersed infinitely thin square platelet system. For an ideal gas, 
\begin{equation}\label{eqn:Ci}
C_{0, ideal}^{-1} = \frac{V^{N}}{N!\Lambda^{3N}},
\end{equation}
where $\Lambda =\frac{h}{\sqrt{2\pi m k_{B}T}}$. Here, $m$ is the mass of a gas particle in the ideal gas. $\Lambda$ is known as the thermal de Broglie wavelength. On the other hand, for platelet systems, 
\begin{equation}\label{eqn:Cp}
C_{0,platelet}^{-1}=\frac{(4\pi V)^{N}}{N!\Lambda^{3N}}. 
\end{equation}
Observe that $C_{0}^{-1}$ changes as $V$ changes. 

Since
\begin{equation}\label{eqn:lCp}
\ln C_{0,platelet}^{-1} = \ln V^{N} + \ln(4\pi)^{N}-\ln N! -\ln \Lambda^{3N} \approx N(\ln \frac{4\pi V}{N \Lambda^{3}} +1),
\end{equation}
we have the following expressions:
\begin{equation}\label{eqn:Fp}
F=-\beta^{-1}\ln C_{0,platelet}^{-1} = -\beta^{-1} N(\ln \frac{4\pi V}{N \Lambda^{3}} +1),
\end{equation}

\begin{equation}\label{eqn:Gp}
G=F+pV=F+\beta^{-1} N =  -\beta^{-1} N(\ln \frac{4\pi V}{N \Lambda^{3}}),
\end{equation}

\begin{equation}\label{eqn:mp}
\mu = \frac{G}{N}= -\beta^{-1} (\ln \frac{4\pi V}{N \Lambda^{3}}) = -\beta^{-1} (\ln \frac{4\pi (\frac{V}{D^3})}{N (\frac{\Lambda}{D})^3}),
\end{equation}

\begin{equation}\label{eqn:bmp}
\beta\mu =  -(\ln \frac{4\pi V}{D^3}) +\ln(\frac{N\Lambda^3}{D^3}).
\end{equation}

\end{document}